\documentclass[12pt]{JHEP3}
\usepackage{amsmath,amssymb,amsfonts}
\usepackage[dvips]{graphicx}
\usepackage{epsfig}

\newcommand{\be}{\begin{equation}}
\newcommand{\ee}{\end{equation}}
\newcommand{\ba}{\begin{eqnarray}}
\newcommand{\ea}{\end{eqnarray}}
\newcommand{\bea}{\begin{array}}
\newcommand{\eea}{\end{array}}

\newcommand{\bw}{{\bf w}}
\newcommand{\br}{{\bf r}}

\newcommand{\tr}{{\rm tr}}

\makeatletter \@addtoreset{equation}{section} \makeatother

\preprint{KIAS-P06019}

\title{\Large BPS String Webs in the 6-dim (2,0) Theories \bf }

\author{Ki-Myeong Lee\\
\\
School of Physics, Korea Institute for Advanced Study, Seoul
130-012
KOREA\\
\email{klee@kias.re.kr} }

\author{Ho-Ung Yee \\
\\
School of Physics, Korea Institute for Advanced Study, Seoul
130-012
KOREA\\
\email{ho-ung.yee@kias.re.kr} }

 \abstract{In the Coulomb phase of the 6-dim (2,0) superconformal theories,
the 1/4, 1/8, 1/16 BPS selfdual string webs are argued to exist
such that the spatial $SO(5)$ and internal $SO(5)$ rotations are
correlated. The basic constituents are 1/2 BPS strings and 1/4 BPS
string junctions. One support comes from the existence of  the
similar BPS dyonic monostring webs in 5-dim maximally
supersymmetric gauge theories. Another comes from the study of the
supersymmetry of the intersecting M2 brane stripes terminating on
M5 branes. We also discuss  the related BPS webs in little string
theories and other theories. }

\begin{document}

\section{Introduction and Conclusion}

There are several mysteries in the 6-dim (2,0) superconformal
theories, which were found from the study of M5 or IIA NS 5 branes
world volume theory and IIB string theory on a ALE
singularity~\cite{Witten:1995zh,Witten:1995em,Seiberg:1997ax,Seiberg:1997zk}
and M theory on $AdS_7\times S^4$~\cite{Maldacena:1997re}. While
they are Lorentz invariant, the theory seems to be non-local field
theory with some nonabelian two-form tensor fields of selfdual
field strength and no coupling constant. In addition from the
study of the near extremal black holes, these theories with the
$U(N)$ group seems to have $N^3$ degrees of freedom in large $N$
from the blakc hole calculation~\cite{Klebanov:1996un}, the
conformal anomaly calculation using the AdS/CFT
correspondence~\cite{Henningson:1998gx}, and the flavor anomaly
calculations~\cite{Harvey:1998bx,Intriligator:2000eq}, leading to
a speculation that maybe basic constituents are three-index
objects instead of two-index objects from the adjoint
representation.

In this article  we argue that in the Coulomb phase of the (2,0)
superconformal  theories,  there exist 1/4, 1/8, 1/16 BPS selfdual
string webs which are made of basic 1/2 BPS selfdual two-indexed
strings, and  1/4 BPS planar three-indexed string junctions. The
spatial rotation group SO(5) is tightly correlated to the internal
R-symmetry SO(5). The low energy theory of their compactification
on a circle are maximally susy 5-dim Yang-Mills theories. We find
the BPS equation for the dyonic monostring webs which are
dimensional reduction of the selfdual string web. We also find
many analogous brane webs in little string theories and other
theories.

The $A_N$ group (2,0) theories arise as the low energy theory of
very closely lying $N+1$ parallel M5 branes at the limit where the
gravity is decoupled. The abelian theory is described by an
antisymmetric tensor field of selfdual field strength and 5 scalar
fields and eight Weyl
fermions\cite{Bandos:1997ui,Aganagic:1997zq}. As the field
strength is selfdual, the nonabelian theory is purely quantum
mechanical, and its nature is mysterious. While there are some
attempts~\cite{Aharony:1997an,Arkani-Hamed:2001ie}, we do not know
the theory well enough to write down the BPS equations.  In the
Coulomb phase where the five scalar fields take nonzero
expectation values and so the gauge symmetry is broken
spontaneously to abelian subgroup, we know that there exist 1/2
BPS selfdual strings. Such BPS selfdual strings were found in the
DBI action of a single M5 branes~\cite{Howe:1997ue}. (For a more
recent attempt to understand the selfdual string in SU(2) theory
from the generalization of the Nahm equation, see, for example,
Ref.~\cite{Basu:2004ed,Berman:2006eu}.)

As the (2,0) theory is not directly approachable, we consider two
somewhat complimentary approaches to the theory.  First  is to
study the compactification of the (2,0) theory on the circle,
whose low energy dynamics is the 5-dim maximally supersymmetric
gauge theories\cite{Seiberg:1997ax}.  We find the explicit BPS
equations for the  dyonic monostring webs and study the simplest
junction in the low energy dynamics of two almost parallel
distinct monostrings. Our BPS equations for the monostring webs
which locks the spatial $SO(4)$ and internal $SO(4)$ of $SO(5)_R$
symmetry turn out to be quite similar to the recent work on the
Langlands program~\cite{Kapustin:2006pk}. Especially our
monostring juctions may have some relevance to the operator
product expansion of the `tHooft operators, which indicates how
two monopole creation operators merge to a single monopole
creation operator.

Another is the study of the supersymmetry of the webs of M2 brane
stripes interpolating the M5 branes by using the world volume
supersymetry~\cite{Bergshoeff:1987qx,Bergshoeff:1987cm}, before
the decoupling limit. We will show that the configuration has 1/32
supersymmetry, implying that the webs has 1/16 supersymmetry on M5
branes. The compactification of a circle along the M5 world volume
They also arise as the web of D2 branes and fundamental strings
interpolating parallel D4 branes in type IIA string theory, which
in turn describe the dyonic monostring webs.

A circle compactification of the space transverse to M5 branes leads
the (2,0) little string theories in the decoupling limit, We show
that there exist the BPS webs of selfdual strings and little strings
in the Coulomb phase of these theories. We also point out that there
can be BPS webs in the (1,1) little string theories and other
theories. (For a review of little string theories, see
Refs.~\cite{Aharony:1999ks}.)

Our BPS string webs are  somewhat similar to the  (p,q) string
junctions and webs in type IIb string
theory~\cite{Dasgupta:1997pu,Sen:1997xi}. In the M-theory context,
these sting webs are understood as the holomorphic embedding of M2
branes, which allows 10-dimensional webs with 1/16 supersymmetry.
Our web for the little string theory can be regarded as the
judicious cutting of some of these M2 brane webs by M5 branes
without supersymmetry breaking. Our analysis indicates that M5
world volume are, say, the real part of the complex 5-dim space,
indicating the junctions should always appear on the 5-brane world
space.

Our work is also partially motivated by trying to understand the
1/4 BPS magnetic monopoles in the (2,0) theories. These dyons are
represented by the(p,q) string junctions ending on D3
branes~\cite{Bergman:1997yw,Lee:1998nv}. Their configuration has
been also studied in the M theory context with $T^2$
compactification of the M5 world volume~\cite{Sasakura:1998cx}.
However in this work, the self-dual string webs is not obvious at
all.

Our consideration of the BPS webs of branes interpolating higher
dimensional branes can appear many other places. One can consider
the 1/4 BPS webs of D5 branes interpolating D7 branes. The field
theory on D5 branes is the N=1 4-dim supersymmetric Yang-Mills
theories. These field theories may be interesting as the 5-dim
field theories on (p,q) 5 brane webs~\cite{Aharony:1997bh}.

Our investigation shows that there exist new classs of BPS webs of
strings and branes in field theories and M-theories. We expect
that there are a lot of degeneracy in the web configuration for a
given boundary condition, whose counting may have relevance with
the blackhole entropy. Our work also indicates that the fields
appear in (2,0) conformal theories belong to the adjoint
representation of the gauge group. In the Coulomb phase the
objects of three group index appear as the string junctions. We
hope that our work is a step forward to understand the nonabelian
nature of the (2,0) conformal theories.

The plan of this work is as follows. In Sec.2, we study the BPS
equations and the webs of dyonic monostrings in the 5-dim
maximally supersymmetric gauge theories. In Sec.3, the
supersymmetry of the M2 brane junctions interpolating M5 branes
are studied. In Sec.4, we make a circle compactification of the
transverse space of the M5 brane and study brane webs in little
string theories. We also study the webs in more general context.

\section{Dyonic Monostring Webs}

The (2,0) theories after the compactification on a circle can be
described in the low energy by the 5-dim  maximally supersymmetric
Yang-Mills theories. This 5-dim theory is not renormalizable and is
completed in the ultraviolet region by the compactified (2,0)
theories.  The size of the compact circle is proportional to the
Yang-Mills coupling constant $e^2/4\pi$, and the (2,0) theory is the
strong-coupling limit of the 5-dim theory. The Kaluza-Klein modes on
the circle appear as instantons, the selfdual strings wrapping the
circle appear as the W-bosons, and the straight strings appear as
the magnetic monostring. They are all 1/2 BPS objects.

With the five scalar field $\phi_A, A=1,2,3,4,5$, the bosonic
Lagrangian is
\be {\cal L}=-\frac{1}{2e^2}\bigg\{  \tr\, F_{MN}F^{MN}+ 2 \tr \,
D_M\phi_A D^M\phi_A+  [\phi_A,\phi_B]^2 \biggr\} \, , \ee
where $D_M\phi_A=\partial_M\phi_A -i[A_M,\phi_A]$ with
$M,N=0,1,2,3,4$. The Gauss law is
\be D_a F_{a0} +i[\phi_a,D_0\phi_a]+i[\phi_5,D_0\phi_5]=0 \, . \ee
We introduce the notation $a=1,2,3,4$, which will be used to relate
the $SO(4)$ spatial symmetry to the $SO(4)$ part of $SO(5)$ internal
space. The conserved energy  is
\ba H &=& \frac{1}{e^2} \int d^4x\, \tr \biggl\{ \sum_a  F_{a0}^2 +
\sum_{a\neq b} \frac{1}{2}F_{ab}^2 + \sum_a (D_0\phi_a)^2  +
\sum_{a,b} (D_a\phi_b)^2 \biggr. \nonumber \\
& & \;\;\;\; \biggl.+ (D_0\phi_5)^2 + \sum_a (D_a\phi_5)^2 -\sum_a
[\phi_5,\phi_a]^2 -\frac{1}{2} \sum_{a\neq b}[\phi_a,\phi_b]^2
\biggr\} \, . \ea
By using the gauss law, we rewrite the energy as
\ba H &=& \frac{1}{e^2} \int d^4x \, \tr \biggl\{\sum_a
(F_{a0}-D_a\phi_5)^2 + (D_0\phi_5)^2 + \sum_a (D_0\phi_a
+i[\phi_5,\phi_a])^2 \biggr. \nonumber \\
& & \;\;\;\;\;\;\; + \frac{1}{2} \sum_{a\neq b}\bigg( F_{ab}
-\epsilon_{abcd}D_c\phi_d
+i[\phi_a,\phi_b]\biggr)^2 + \biggl(\sum_a D_a\phi_a\biggr)^2 \biggr\} \nonumber \\
& & + Q_M + Q_E \, , \ea
where the topological magnetic and electric energies are,
respectively,
\ba && Q_M = \frac{1}{e^2}\int d^4x\; \partial_a \; \tr \; \biggl\{
\epsilon_{abcd} \bigl(\phi_b F_{cd} +\frac{i}{3}\phi_b
[\phi_c,\phi_d] \bigr) +(\phi_b D_b\phi_a - \phi_a
D_b\phi_b)\biggr\} ,  \nonumber
\\
&&  Q_E= \frac{1}{e^2} \int d^4x\;
\partial_a \, \tr(F_{a0}\phi_5)\, . \ea
The energy bound is saturated by the  BPS configurations which are
static in time by a gauge choice and satisfy $A_0=\phi_5$ and
\ba && F_{ab}-\epsilon_{abcd}D_c\phi_d +i[\phi_a,\phi_b]=0 \,
,\,\,  D_a\phi_a=0\, ,\nonumber \\
&&  D_a^2\phi_5-[\phi_a,[\phi_a,\phi_5]]=0 \, .\label{selfdualeq}
\ea
The last equation originates from the Gauss law, and decides the
$\phi_5$ in the given $A_a, \phi_a$ solution of the first part of
the BPS equations.  The first seven equations have also appeared
recently as Eq.(3.29) of Ref.~\cite{Kapustin:2006pk} which
discusses the geometric  Langlands program. One can show the above
equations imply that the BPS configurations satisfy the field
equations. While the instanton number does not appear in the BPS
energy, one can show easily that the instanton number remains
topological even after one uses the BPS equations. Once the
solution is found, one can make the spatial O(4) and internal O(4)
transformations independently. After using the BPS equations, the
magnetic and electric energies become, respectively,
\ba && Q_M=\frac{1}{e^2}\int d^4x \,\partial_a \left\{
\partial_b (\delta_{ab}\tr\phi_c^2-\tr\phi_a\phi_b)
-\frac{2i}{3} \epsilon_{abcd} \tr
\phi_b[\phi_c,\phi_d]\right\}\, , \\
&& Q_E= \frac{1}{e^2}\int d^4x \, \partial_a^2\, \tr \phi_5^2 \, .
\ea

To consider the supersymmetry, we use the 10-dim real gamma
matrices, and choose our spinors to be Majorana and Weyl. Under the
supersymmetry, the  gaugino field transform
$\delta\lambda=\frac{1}{2} F_{PQ}\Gamma^{PQ}$, with $P,Q=0,1,...9$
becomes
\ba \delta \lambda
 &=& \frac{1}{2}\Gamma^{ab} \biggl(F_{ab}
+i[\phi_a,\phi_b]\Gamma^{ab,4+a,4+b} -\epsilon_{abcd}D_c\phi_d
\Gamma^{abc,4+d} \biggr)\epsilon \nonumber \\
& & + \Gamma^{15}\biggl(D_1\phi_1 + D_2\phi_2 \Gamma^{1256}+
D_3\phi_3\Gamma^{1357}+D_4\phi_4\Gamma^{1458} \biggr)\epsilon
\nonumber \\
& & +\Gamma^{a0}(F_{a0}-D_a\phi_5 \Gamma^{09} ) + \Gamma^{0,4+a}
(D_0\phi_a + i[\phi_5,\phi_a]\Gamma^{09})  \epsilon +
D_0\phi_5\Gamma^{09}\epsilon\, . \ea
On the supersymmetric parameter $\epsilon$, we can impost the
mutually consistent and independent four conditions
\be \Gamma^{2345}\epsilon=-\epsilon,\;
\Gamma^{1346}\epsilon=\epsilon, \;
\Gamma^{1247}\epsilon=-\epsilon,\; \Gamma^{1238}\epsilon=\epsilon \,
. \ee
Then one can show easily that  $\delta \lambda=0$ if the self-dual
equations (\ref{selfdualeq}) are satisfied. This shows that our
configuration is 1/16 BPS. The spatial 1/16 BPS equation in 8+1
Yang-Mills has been obtained in Ref.~\cite{Bak:2002aq}, which can be
dimensionally reduced to be the first part of the BPS
equations~(\ref{selfdualeq}). If we have chosen different sign
requirement for the $\epsilon$, we still get the BPS equations, but
the relation between the spatial rotation group $SO(4)$ and the
internal $SO(4)$ group would  not manifest.

Recall that this 5-dim theory with $SU(N)$ gauge group arises from
the low energy dynamics of parallel D4 branes  very close to each
other in the type IIa string theory. At the ground state
$[\phi_A,\phi_B]=0$ and so their expectation values upon
multiplication of $\ell_s^2$ denote the position of D4 branes in
the transverse 5-dim space. Assuming the D4 world volume to be on
the $01234$ space of the target space, the higgs fields
expectation value  $\phi_a$ denote  the position of D4 branes on
the transverse $x^{a+4}$ space. Our BPS magnetic monostring webs
are made of webs of D2 brane stripes interpolating D4 branes.

The simplest case arises with only one scalar field, say $\phi_1$,
has nonzero expectation value, and so D4 branes lie along a line. We
ignore the electric charge part for a while. The BPS equations
become
\be F_{23}+D_4\phi_1=0,\; F_{34}+D_2\phi_1=0,\; F_{42}+D_3\phi_1=0 ,
\; D_1\phi_1=0\, , \ee
which are the BPS equation for magnetic monostrings along $x^1$
direction. For the SU(2) gauge group with the scalar field
expectation value $\phi={\rm diag}(v,-v)/2$, the solution for unit
magnetic charge would be simply the BPS magnetic monopole solution
whose radial variable is given by the radial variable transverse to
the monostring. The string tension of a monostring of unit magnetic
flux would be $4\pi v/e^2$.

The next simple case involves $\phi_1,\phi_2$ only, and so the D4
branes lie on a plane. The BPS equations (\ref{selfdualeq}) become
\ba &&  F_{12}+i[\phi_1,\phi_2]=0, \,
F_{34}-D_1\phi_2+D_2\phi_1=0,\,
F_{13}-D_4\phi_2=0, \, F_{14}+D_3\phi_2=0, \nonumber \\
&&  F_{23}+D_4\phi_1=0, \, F_{24}-D_3\phi_1=0,\,
D_1\phi_1+D_2\phi_2=0 . \label{sd4} \ea
With the $SU(2)$ gauge group with the Higgs expectation value
\be \phi_1={\rm diag}(v,-v)\cos\theta/2\, ,\;\; \phi_2={\rm
diag}(v,-v)\sin\theta/2\, , \ee
the magnetic monostring solution of the above equation (\ref{sd4})
is now along the line which is rotated on the 12 plane by the angle
$\theta$ from the $x^1$ axis. In the $SU(2)$ gauge theory, we  can
consider all scalar fields $\phi_a$ taking expectation value. The
solution of our BPS equations (\ref{selfdualeq}) for the $SU(2)$
case implies that the direction of  BPS monostrings  in 1234 space
is identical to the line connecting two points in  5678 space given
by the $\phi_a$ expectation value. In terms of D2 branes connecting
two D4 branes, the direction of D2 branes on the D4 world volume is
identical to the direction of D2 branes on the transverse space of
D4 branes. This locking of the internal rotational group in the
world volume  and the external rotation in the transverse space
would be again manifest from the study of the supersymmetry of M2
brane stripes interpolating parallel M5 branes.

With the $SU(3)$ gauge group, the  Higgs expectation value
\be \phi_1={\rm diag}(\mu_1,0,-\mu_2),  \;\; \phi_2 = {\rm diag}(0,
c, 0)/2 \ee
denote the position of the three D4 branes on  the transverse
$x^5,x^6$ plane upon the multiplication of $\ell_s^2$. There can be
three magnetic monostrings, represented by the D2 branes connecting
two of three D4 branes.  We order three D4 branes by their positions
are $(\mu_1,0), (-\mu_2,0), (0,c)$ on the 56 plane. We put
$\mu_1,\mu_2,c$ to be positive for the convenience. The tension of
three magnetic monostrings become
\be  T_{12}=\frac{4\pi}{e^2} (\mu_1+\mu_2)h \; ,\;\;
T_{13}=\frac{4\pi}{e^2}\sqrt{\mu_1^2+c^2}\;  ,\;\;
T_{23}=\frac{4\pi}{e^2}\sqrt{\mu_2^2+c^2}\; .\ee
Each magnetic monostrings as the solution of the BPS equation
(\ref{sd4}) would have tilting on 12 plane determined by the Higgs
expectation value. Thus $\tan\theta_{12}=0,
\tan\theta_{13}=-c/\mu_1$, and $\tan\theta_{23}=c/\mu_2$. We move
these three monostrings keeping their directions and so they meet at
a point in 12 plane. Then we make a junction of three monostrings.
One can see easily that the tension valance condition, which are
$T_{12} \cos\theta_{12}= T_{13}\cos_{13}+T_{23}\cos_{23}$ and
$T_{12}\sin\theta_{12}=T_{13}\sin\theta_{13}+T_{23}\sin\theta_{23}$,
are satisfied.

We have somewhat different tools to  check this fact. Let us
recall the low energy dynamics of two distinct monopoles, which is
described by the Taub-NUT
metric~\cite{Gauntlett:1996cw,Lee:1996if}. With two
non-proportional scalar field expectation value, the nonlinear
sigma model of the Taub-NUT metric would be modified by the
nonlinear sigma model with potential for the moduli
coordinates~\cite{Tong:1999mg,Bak:1999da}. One expects naively
that the same Lagrangian goes over to the low energy effective
Lagrangian for two parallel  magnetic monostrings, where the
typical wave length for the fluctuations is much larger than the
separation between strings. Thus we consider the above $SU(3)$
model with the parameter $c$ of $\phi_2$ expectation value being
very small.

Thus we consider two parallel monostrings  of tension $T_{13}$ and
$T_{23}$, which are combined to a single string of tension $T_{12}$.
Suppose they are lying along the $x^1=x$ axis and their relative
position $\br=(x^2,x^3,x^4)$ is given along the transverse
direction.  As the relative position and phase are now the functions
of time and world sheet coordinate $x=x^1$, the relative moduli
space dynamics is described by  1+ 1 dimensional nonlinear sigma
model,
\be \! {\cal L}_{1+1} =- \frac{g\mu}{2} \biggl( (1+
\frac{1}{2\mu|\br|}) (\partial_\alpha \br )^2
+\frac{1}{4\mu^2(1+\frac{1}{2\mu |\br|} ) }(\partial_\alpha \psi
+\bw\cdot
\partial_\alpha \br )^2\biggr) - \frac{g}{2\mu} \frac{c^2}{1+\frac{1}{1+2\mu|\br|} }\, , \ee
where  the relative tension $\mu=\mu_1\mu_2/(\mu_1+\mu_2)$ and
$g=4\pi/e^2$. We are considering the limit where $c$ is much smaller
than $\mu_1,\mu_2$.

We expect that the composite monopole string of tension
$T_{12}=g(\mu_1+\mu_2)$ lying along $x^1$ direction gets split to
two monopole strings of tension $T_{13}$ and $T_{13}$ on $x^1,x^2$
plane.   To find the configuration, let us assume that two strings
are separated along the $\br(x=x^1) = ( y=x^2,0,0)(x)$. The relevant
energy becomes
\be E_{string}=\int dx\left\{ \frac{g\mu}{2}\biggl(1+
\frac{1}{2\mu|y|}\biggr) (\partial_x y)^2
+\frac{g}{2\mu}\frac{c^2}{1+ \frac{1}{2\mu |y|}} \right\}\, . \ee
We can reexpress it as
\be E_{string} = \int dx\;\; \frac{g\mu}{2}\biggl(1+\frac{1}{2\mu
|y|}\biggr) \biggl( \partial_x y \mp
\frac{c}{\mu(1+\frac{1}{2\mu|y|})} \biggr)^2 \pm  g c \int dx
\partial_x y\, . \label{string1}\ee
The solution $y(x)$ for the BPS equation is implicitly given by
\be 2\mu y e^{2\mu y} = e^{2cx} \, ,\ee
for the junction point at $x=y=0$.  The separation parameter $y(x)$
goes to zero exponentially fast when $x\rightarrow -\infty$ and
grows linearly with $x$ with slope $c/\mu$ for large $x$. For two
string branches emerging from the $1-3$ string would be described by
the positions $y_1(x)$ and $y_2(x)$. The center of mass position
remain to be zero and so $\mu_1 y_1 + \mu_2 y_2=0$. As the relative
position $y=y_1-y_2$, we see that two string positions diverge like
$y_1(x)= cx/\mu_1$ and $y_2(x)=cx/\mu_2$. Not only the angles of
these two strings are exactly what expected from the tension
valence, the boundary term (\ref{string1}) for the energy matches
the total sum of energies which takes the slanting of the strings
into account. Thus the moduli space dynamics of two distinct
monostrings gives the junction configuration.

\section{Webs of  M2 brane stripes terminating at M5 branes}

To argue that there exist 1/4, 1/8, 1/16 BPS webs of selfdual
strings  in the (2,0) conformal theories, we return to the M theory
origin of these theories. For simplicity, we consider the $SU(N)$
gauge group arises from the  $N$ M5 branes. The webs of selfdual
strings would appear as the webs of M2 brane stripes interpolating
parallel M5 branes in the M theory. The consideration of the
preserved supersymmetry of the web configuration is more less
similar to the monostring web case.

Let us start with the $SU(2)$ case with the two parallel M5 branes
with the world voluem on the 012345 space and take arbitrary
positions on $6789\bar{10}$ space. They are 1/2 BPS configuration of
the M theory as the spinor parameter $\epsilon$ for the kappa
symmetry of these M5 branes satisfies in the 11-dim gamma matrices
\be \Gamma^{012345}\epsilon=\epsilon\, . \label{m5susy} \ee
It is possible for M2 branes to end on M5 branes. They are BPS as
long as M2 branes and M5 branes are planar and M2 branes are
orthogonal to M5 branes~\cite{Strominger:1995ac}. The preserved
supersymmetry is identical to that for the intersecting M5 and M2
branes without termination.

Let say first two M5 branes are separated along the $x^{10}$
coordinate only.  We now consider to put M2 brane stripes connecting
these M5 branes, where all M2 branes are lying on $5\bar{10}$ plane.
The spinor for the kappa-symmetry on the M2 brane connecting these
two M5 branes would satisfy
\be \Gamma^{05\bar{10}}\epsilon=\epsilon\, . \label{m2susy1}\ee
The above condition is compatible with the condition (\ref{m5susy})
on M5 and so the configuration is  1/4 BPS or has 8 supersymmetries.
Many such parallel  M2 brane stripes  can be inserted maybe with
different  $1234$ space. It describes the parallel  selfdual strings
in the (2,0) conformal theories.

Let us now imagine adding a M5 brane whose location is on the
$9\bar{10}$ plane. Now we have three M5 branes in the $9\bar{10}$
plane, and so three possible M2 branes connecting them. This is
exactly similar to the monopole junction case for the $SU(3)$ group.
For two M5 branes with the connecting line making an angle $\alpha$
with $x^{10}$ axis, we  rotate the M2 brane stripe interpolating
them in both the $4,5$ plane and the $9\bar{10}$ plane by the same
angle. The spinor parameter for the  $\kappa$ symmetry of the second
M2 brane is preserved if
\be
\Gamma^{09\bar{10}}e^{\alpha(\Gamma^{45}+\Gamma^{9\bar{10}})}\epsilon=\epsilon\,
, \ee
which is true for any angle $\alpha$ if the condition
(\ref{m2susy1}) and the following equation holds:
\be \Gamma^{049}\epsilon=\epsilon\, . \label{m2susy2}\ee
Thus we can now have a BPS configuraiton of  three M2 brane stripes
connecting three parallel M5 branes. M2 branes are intersecting each
other in angles~\cite{Berkooz:1996km,Papadopoulos:1996uq}, remaining
BPS. If they are not terminating on M5 branes, two intersecting M2
branes are meeting at a point in $459\bar{10}$ space. But as they
are terminating on M5 branes and so have effective
thickness~\cite{Callan:1997kz}, and so their intersection region
becomes nonabelian and fuzzy. When these three M2 branes terminating
on M5 branes are meeting at the same point or fuzzy region, they can
make a BPS planar junction instead of continuing as the straight
stripes. While it is very nontrivial configurations, the tension
balance and the five dimensional monopole string junction support
this possibility.

This identification of the two rotation groups $SO(5)_S\times
SO(5)_T$  of M2 brane stripes in the $12345$ space and in the
transverse $6789\bar{10}$ is essential for the supersymmetry to be
maintained. Thus one  can add more M5 branes with arbitrary location
in $6789\bar{10}$ space and M2 branes stripes in similar manner and
the total configuration  will remain supersymmetric if the following
independent and mutually consistent conditions are satisfie,
\be \Gamma^{012345}\epsilon=\epsilon\, , \;\;
\Gamma^{027}\epsilon=\epsilon\, , \;\;
\Gamma^{038}\epsilon=\epsilon\, ,\;\;
\Gamma^{049}\epsilon=\epsilon\, , \;\;
\Gamma^{01\bar{10}}\epsilon=\epsilon \, . \label{m2susy3}\ee
{}From the condition that $\Gamma^{012\cdots 9\bar{10}}=1$ for
11-dim gamma matrices, the imposing conditions (\ref{m2susy3}) on M2
branes implies that $ \Gamma^{016}\epsilon=\epsilon $. Thus we would
have the 1/32 BPS  web configurations of M2 brane stripe connecting
parallel M5 branes.

\section{Little String Theories and Other Theories}

One can imagine compactifying some spatial direction of the above
M5-M2 brane configuration. Let us first start with compactifying
$x^{\bar{10}}$, which does not interfere with the above
supersymmetry argument. For two M5 branes lying on $9\bar{10}$
plane, one can have now two kinds of BPS M2 brane stripes connecting
these M5 branes. One wrapping clockwise and another wrapping
counter-clockwise such that  such that the total winding number of
$x^10$ is unity.  When two M5 branes do not line on the same $x^9$
coordinates, these two kinds of M2 branes stripes are attracted to
each other trying to form a M2 brane which winds $x^10$ once. (This
is similar to the case where calorons appears a composite of
magnetic monopoles~\cite{Lee:1997vp}.) Thus, naturally they can form
a junction. As the compactifying the $x^{10}$ is like introducing
infinitely number of $M5$ branes and so there can be more
complicated junctions with more winding numbers. Adding more M5
branes with different transverse directions allow webs of less
supersymmetry and non-planar as in the previous case. With more M5
branes, one can now build the BPS webs of M2 brane stripes which
interpolate M5 branes and the whole M2 branes which wrap $x^{10}$
integer times.

In the little string theory limit where the gravity decouples in the
above picture, a BPS little string of $(2,0)$ theory which arises as
M2 brane winding the $x^{10}$ once, appears as a singlet under the
gauge symmetry. The above junction would appear on the SU(2) (2,0)
little string theory as the planar junction of a little string and
two complimentary selfdual strings on $45$ plane. Similary, the webs
would become the BPS webs of little strings and selfdual strings in
the (2,0) little string theories.

Let us now consider  compactifying the world sheet volume of two M5
branes, which have two different  $x^{\bar{10}}$ positions. Let us
try to compactify $x^4,x^5$ direction of M5 direction. The M2 brane
sheets lying along $x^4,x^5$ and $x^9,x^{\bar{10}}$ direction and
connecting two M5 branes would appear (p,q) dyons on the 3+1 dim
field theory along $0,1,2,3$ direction of wrapped M5 branes.  We
have another BPS (-p,-q) dyons by warping another $M2$ stripes which
wrap $x^{\bar{10}}$ complimentary way and connect the same two M5
branes. In general these two distinct BPS M2 stripes have different
tilting angle on $x^4,x^5$ plane. At least for the special case of
the torus shape and M5 positions, one can wind these two M2 brane
stripes opposite way in 4-5 torus so that their charge cancel each
other. We suspect that in generic case there would be web
configurations of these stripes which would have zero charge.

Without costing the divergent energy nonzero charge, one can now
further compactify an additional M5 world sheet direction, say
$x^3$. Then if there is additional M5 branes whose location is not
aligned in $x^8$ direction, one can have nontrivial wrap on 5-3
torus, again making the total charge zero. Similarly one can
compactify $x^2$ direction with more M5 branes not aligned along
$x^7$ direction. This  configuration would be now 1/32 BPS as the
supersymmetry (\ref{m2susy3}) indicates. Now we have effective 1+1
world sheet theory in Coulomb phase of M5 branes wrapping $T^4$ of
$T^5$ space, on which there are  a lump of tangled web of M2 stipes.

Once one compactify $x^1$ direction similar way, all M5 branes are
pulled together in the transverse direction and tends to align along
the compact direction $x^6$. It would be interesting whether still
there is 1/32 BPS configurations of lumps in $R^{4+1}$ space with
$T^6$ compactification, with M5 branes wrapping $T^5$. They would be
characterized by many charges with large degeneracy, and could  be
represented by 1/32 BPS extremal blackholes.

In somewhat different extension of the above web configurations can
be found. For the little string theory of $(1,1)$ theory in 6-dim,
the low energy field theory is 6-dim (1,1) Yang-Mills theory and the
monopole sheet (as generalized from 5-dim) and instanton strings are
low energy description of the membrane and little strings,
respectively. Instanton strings can terminate at the monopole
membrane. Monopole membranes and instanton strings can form
complicated BPS webs whose BPS equation in the Yang-Mills language
is similar to the monopole string webs. Only change is to uplift one
scalar field, say $\phi_1$ for $x^5$ directional position to a gauge
field $A_5$, and so the BPS equations get modified as
$D_a\phi_1\rightarrow F_{a5}$, and $-i[\phi_1,\phi_b]\rightarrow
D_5\phi_b$ for $b=2,3,4,5$.  The little $(1,1)$ string theory can be
lifted to the little $m$ theory in 7 dim~\cite{Losev:1997hx}. Thus
the above membrane string web in the (1,1) little string theory  can
be lifted to the little $m$ theory. Similarly the web in the (2,0)
little string theory can be studied in the 8-dim little $f$ theory.

\noindent{\bf Acknowlegement}

This work is supported in part by KOSEF Grant
R010-2003-000-10391-0 (K.L.,H.U.Y.), KOSEF SRC Program through
CQUeST at Sogang Univ. (K.L.), KRF Grant No. KRF-2005-070-C00030
(K.L.). We appreciate useful discussions with Dongsu Bak, M.
Henningson, A. Kapustin, Seok Kim, Soo-jong Rey, A. Strominger,
and Piljin Yi.


\begin{thebibliography}{12345}



\bibitem{Witten:1995zh}
  E.~Witten,
  ``Some comments on string dynamics,''
  arXiv:hep-th/9507121.

\bibitem{Witten:1995em}
  E.~Witten,
  ``Five-branes and M-theory on an orbifold,''
  Nucl.\ Phys.\ B {\bf 463}, 383 (1996)
  [arXiv:hep-th/9512219].

\bibitem{Seiberg:1997ax}
  N.~Seiberg,
  ``Notes on theories with 16 supercharges,''
  Nucl.\ Phys.\ Proc.\ Suppl.\  {\bf 67}, 158 (1998)
  [arXiv:hep-th/9705117].


\bibitem{Seiberg:1997zk}
  N.~Seiberg,
  ``New theories in six dimensions and matrix description of M-theory on  T**5
  and T**5/Z(2),''
  Phys.\ Lett.\ B {\bf 408}, 98 (1997)
  [arXiv:hep-th/9705221].


\bibitem{Maldacena:1997re}
  J.~M.~Maldacena,
  ``The large N limit of superconformal field theories and supergravity,''
  Adv.\ Theor.\ Math.\ Phys.\  {\bf 2} (1998) 231
  [Int.\ J.\ Theor.\ Phys.\  {\bf 38} (1999) 1113]
  [arXiv:hep-th/9711200].



\bibitem{Klebanov:1996un}
  I.~R.~Klebanov and A.~A.~Tseytlin,
  ``Entropy of Near-Extremal Black p-branes,''
  Nucl.\ Phys.\ B {\bf 475}, 164 (1996)
  [arXiv:hep-th/9604089].



\bibitem{Henningson:1998gx}
  M.~Henningson and K.~Skenderis,
  ``The holographic Weyl anomaly,''
  JHEP {\bf 9807} (1998) 023
  [arXiv:hep-th/9806087].



\bibitem{Harvey:1998bx}
  J.~A.~Harvey, R.~Minasian and G.~W.~Moore,
  ``Non-abelian tensor-multiplet anomalies,''
  JHEP {\bf 9809} (1998) 004
  [arXiv:hep-th/9808060].

\bibitem{Intriligator:2000eq}
  K.~A.~Intriligator,
  ``Anomaly matching and a Hopf-Wess-Zumino term in 6d, N = (2,0) field
  Nucl.\ Phys.\ B {\bf 581} (2000) 257
  [arXiv:hep-th/0001205].


\bibitem{Bandos:1997ui}
  I.~A.~Bandos, K.~Lechner, A.~Nurmagambetov, P.~Pasti, D.~P.~Sorokin and M.~Tonin,
  ``Covariant action for the super-five-brane of M-theory,''
  Phys.\ Rev.\ Lett.\  {\bf 78}, 4332 (1997)
  [arXiv:hep-th/9701149].

\bibitem{Aganagic:1997zq}
  M.~Aganagic, J.~Park, C.~Popescu and J.~H.~Schwarz,
  ``World-volume action of the M-theory five-brane,''
  Nucl.\ Phys.\ B {\bf 496}, 191 (1997)
  [arXiv:hep-th/9701166].



\bibitem{Aharony:1997an}
  O.~Aharony, M.~Berkooz and N.~Seiberg,
  ``Light-cone description of (2,0) superconformal theories in six
  dimensions,''
  Adv.\ Theor.\ Math.\ Phys.\  {\bf 2}, 119 (1998)
  [arXiv:hep-th/9712117].


\bibitem{Arkani-Hamed:2001ie}
  N.~Arkani-Hamed, A.~G.~Cohen, D.~B.~Kaplan, A.~Karch and L.~Motl,
  ``Deconstructing (2,0) and little string theories,''
  JHEP {\bf 0301}, 083 (2003)
  [arXiv:hep-th/0110146].



\bibitem{Howe:1997ue}
  P.~S.~Howe, N.~D.~Lambert and P.~C.~West,
  ``The self-dual string soliton,''
  Nucl.\ Phys.\ B {\bf 515} (1998) 203
  [arXiv:hep-th/9709014].


\bibitem{Basu:2004ed}
  A.~Basu and J.~A.~Harvey,
  ``The M2-M5 brane system and a generalized Nahm's equation,''
  Nucl.\ Phys.\ B {\bf 713} (2005) 136
  [arXiv:hep-th/0412310].

\bibitem{Berman:2006eu}
  D.~S.~Berman and N.~B.~Copland,
  ``A note on the M2-M5 brane system and fuzzy spheres,''
  arXiv:hep-th/0605086.



\bibitem{Kapustin:2006pk}
  A.~Kapustin and E.~Witten,
  ``Electric-magnetic duality and the geometric Langlands program,''
  arXiv:hep-th/0604151.


\bibitem{Bergshoeff:1987qx}
  E.~Bergshoeff, E.~Sezgin and P.~K.~Townsend,
  ``Properties Of The Eleven-Dimensional Super Membrane Theory,''
  Annals Phys.\  {\bf 185}, 330 (1988).

\bibitem{Bergshoeff:1987cm}
  E.~Bergshoeff, E.~Sezgin and P.~K.~Townsend,
  ``Supermembranes And Eleven-Dimensional Supergravity,''
  Phys.\ Lett.\ B {\bf 189}, 75 (1987).




\bibitem{Aharony:1999ks}
  O.~Aharony,
  ``A brief review of 'little string theories',''
  Class.\ Quant.\ Grav.\  {\bf 17}, 929 (2000)
  [arXiv:hep-th/9911147].








\bibitem{Dasgupta:1997pu}
  K.~Dasgupta and S.~Mukhi,
  ``BPS nature of 3-string junctions,''
  Phys.\ Lett.\ B {\bf 423}, 261 (1998)
  [arXiv:hep-th/9711094].


\bibitem{Sen:1997xi}
  A.~Sen,
  ``String network,''
  JHEP {\bf 9803}, 005 (1998)
  [arXiv:hep-th/9711130].



\bibitem{Krogh:1997dx}
  M.~Krogh and S.~M.~Lee,
  ``String network from M-theory,''
  Nucl.\ Phys.\ B {\bf 516} (1998) 241
  [arXiv:hep-th/9712050].

\bibitem{Matsuo:1997jw}
  Y.~Matsuo and K.~Okuyama,
  ``BPS condition of string junction from M theory,''
  Phys.\ Lett.\ B {\bf 426} (1998) 294
  [arXiv:hep-th/9712070].



\bibitem{Bergman:1997yw}
  O.~Bergman,
  ``Three-pronged strings and 1/4 BPS states in N = 4 super-Yang-Mills
  Nucl.\ Phys.\ B {\bf 525} (1998) 104
  [arXiv:hep-th/9712211].

\bibitem{Lee:1998nv}
  K.~M.~Lee and P.~Yi,
  ``Dyons in N = 4 supersymmetric theories and three-pronged strings,''
  Phys.\ Rev.\ D {\bf 58} (1998) 066005
  [arXiv:hep-th/9804174].


\bibitem{Sasakura:1998cx}
  N.~Sasakura and S.~Sugimoto,
  ``M-theory description of 1/4 BPS states in N = 4 supersymmetric  Yang-Mills
  Prog.\ Theor.\ Phys.\  {\bf 101} (1999) 749
  [arXiv:hep-th/9811087].




\bibitem{Aharony:1997bh}
  O.~Aharony, A.~Hanany and B.~Kol,
  ``Webs of (p,q) 5-branes, five dimensional field theories and grid
  JHEP {\bf 9801} (1998) 002
  [arXiv:hep-th/9710116].



\bibitem{Bak:2002aq}
  D.~s.~Bak, K.~M.~Lee and J.~H.~Park,
  ``BPS equations in six and eight dimensions,''
  Phys.\ Rev.\ D {\bf 66} (2002) 025021
  [arXiv:hep-th/0204221].


\bibitem{Gauntlett:1996cw}
  J.~P.~Gauntlett and D.~A.~Lowe,
  ``Dyons and S-Duality in N=4 Supersymmetric Gauge Theory,''
  Nucl.\ Phys.\ B {\bf 472} (1996) 194
  [arXiv:hep-th/9601085].

\bibitem{Lee:1996if}
  K.~M.~Lee, E.~J.~Weinberg and P.~Yi,
  ``Electromagnetic Duality and $SU(3)$ Monopoles,''
  Phys.\ Lett.\ B {\bf 376} (1996) 97
  [arXiv:hep-th/9601097].


\bibitem{Tong:1999mg}
  D.~Tong,
  ``A note on 1/4-BPS states,''
  Phys.\ Lett.\ B {\bf 460} (1999) 295
  [arXiv:hep-th/9902005].

\bibitem{Bak:1999da}
  D.~Bak, C.~k.~Lee, K.~M.~Lee and P.~Yi,
  ``Low energy dynamics for 1/4 BPS dyons,''
  Phys.\ Rev.\ D {\bf 61}, 025001 (2000)
  [arXiv:hep-th/9906119].




\bibitem{Strominger:1995ac}
  A.~Strominger,
  ``Open p-branes,''
  Phys.\ Lett.\ B {\bf 383}, 44 (1996)
  [arXiv:hep-th/9512059].






\bibitem{Berkooz:1996km}
  M.~Berkooz, M.~R.~Douglas and R.~G.~Leigh,
  ``Branes intersecting at angles,''
  Nucl.\ Phys.\ B {\bf 480}, 265 (1996)
  [arXiv:hep-th/9606139].


\bibitem{Papadopoulos:1996uq}
  G.~Papadopoulos and P.~K.~Townsend,
  ``Intersecting M-branes,''
  Phys.\ Lett.\ B {\bf 380} (1996) 273
  [arXiv:hep-th/9603087].





\bibitem{Callan:1997kz}
  C.~G.~.~Callan and J.~M.~Maldacena,
  ``Brane dynamics from the Born-Infeld action,''
  Nucl.\ Phys.\ B {\bf 513}, 198 (1998)
  [arXiv:hep-th/9708147].






\bibitem{Lee:1997vp}
  K.~M.~Lee and P.~Yi,
  ``Monopoles and instantons on partially compactified D-branes,''
  Phys.\ Rev.\ D {\bf 56}, 3711 (1997)
  [arXiv:hep-th/9702107].







\bibitem{Losev:1997hx}
  A.~Losev, G.~W.~Moore and S.~L.~Shatashvili,
  ``M \& m's,''
  Nucl.\ Phys.\ B {\bf 522} (1998) 105
  [arXiv:hep-th/9707250].



\end{thebibliography}
\end{document}